\documentclass[preprint]{aastex}

\def\lea{\mathrel{<\kern-1.0em\lower0.9ex\hbox{$\sim$}}}
\def\gea{\mathrel{>\kern-1.0em\lower0.9ex\hbox{$\sim$}}}

\slugcomment{Accepted by The Astrophysical Journal}

\shorttitle{RR Lyrae Variables in the Local Group Dwarf Galaxy NGC 147}
\shortauthors{Soung-Chul Yang \& Ata Sarajedini}

\begin{document}

\title{RR Lyrae Variables in the Local Group Dwarf Galaxy NGC 147}

\author{S-C. Yang and Ata Sarajedini}
\affil{Department of Astronomy, University of Florida,
    Gainesville, FL 32611}
\email{sczoo@astro.ufl.edu, ata@astro.ufl.edu}


\begin{abstract}
We investigate the RR Lyrae population in NGC 147, a dwarf satellite galaxy 
of M31 (Andromeda). We used both Thuan-Gunn $g$-band ground-based photometry 
from the literature and Hubble Space Telescope Wide Field Planetary 
Camera 2 archival data in the F555W and F814W passbands to investigate
the pulsation properties of RR Lyrae variable candidates in NGC 147. 
These datasets represent the two extreme cases often found in 
RR Lyrae studies with respect to the phase coverage of the 
observations and the quality of the photometric measurements.
Extensive artificial variable star tests for both cases were performed. 
We conclude that
neither dataset is sufficient to confidently determine the pulsation
properties of the NGC 147 RR Lyraes. 
Thus, while we can assert that NGC 147 contains RR Lyrae
variables, and therefore a population older than $\sim$10 Gyr, it is
not possible at this time to use the pulsation properties of these RR Lyraes
to study other aspects of this old population.
Our results provide a good 
reference for gauging the completeness of RR Lyrae variable detection
in future studies.
  
\end{abstract}


\section{Introduction}
Local Group dwarf galaxies exhibit diversity in their star formation histories 
(Da Costa \& Mould 1998; Grebel 1999). Most of these galaxies show some evidence 
for intermediate age stars and some have even younger populations along with a 
significant amount of gas and dust. The broad red giant branch (RGB) morphology 
shown in their color-magnitude diagrams (CMDs) reflects a wide range of 
metallicities and/or ages, implying that these galaxies have experienced complex 
star formation histories. Despite their different and complex star formation 
histories, RR Lyrae variables are believed to exist in most if not all dwarf galaxies in the 
Local Group. The existence of RR Lyrae (RRL) populations in a given stellar system 
indicates that the system is older that $\sim$10 Gyr. Furthermore, their pulsation 
properties and absolute magnitudes are correlated with their metallicities
(Bono et al. 2003; Bono et al. 2007; Kunder \& Chaboyer 2009). 
Therefore, detailed investigations of the physical properties of RRL 
variable stars are crucial to improve our understanding of the early stages of star 
formation in Local Group dwarf galaxies. 

NGC 147 is one of the dwarf spheroidal (dSph) satellites of the Andromeda 
galaxy (M31). In terms of its morphology, NGC 147 is a typical dwarf elliptical or 
spheroidal galaxy. However, it is distinct from other dwarf galaxies in the Local 
Group, with the possible exception of Tucana (Fraternali et al. 2009),
because NGC 147 is dominated by an old stellar population (Hodge 1989). 
There is a slight indication of intermediate age stars (e.g. extended asymptotic 
giant branch stars) concentrated in the central regions (Han et al. 1997), 
but their signature is weaker than the intermediate age stellar populations found 
in other Local Group dwarf galaxies. Furthermore, NGC 147 appears to
show a complete deficit of gas and dust. 

The presence of an RR Lyrae population in NGC 147 was reported by 
Saha et al. (1990, hereafter S90). They used Thuan-Gunn 
$g$-band ground based photometry 
obtained from the Hale 5m telescope equipped with the 4-shooter CCD system 
to identify RR Lyrae variables in this galaxy. They found 32 RR Lyrae candidates and 
determined the distance modulus of NGC 147 to be $(m-M)_0$= 23.92 based
on their mean apparent $g$-band magnitudes. Their periods and amplitudes were 
estimated by using a prototype of the phase dispersion minimization method 
developed by Lafler \& Kinman (1965), the so-called L-K method. Saha et al. 
(1990) did not present a comparison of the pulsation properties of these RR Lyrae 
candidates with those in other Local Group galaxies. 
One reason for this may be that the large photometric errors ( $>$ 0.15 mag) 
at the level of the horizontal branch (HB) of NGC 147 produced significant
errors in the period determinations - especially for the shorter period 
RR Lyraes (P $<$ 0.4 d). However, the effect of the photometric errors on 
the period determination was not addressed because the L-K method ignores 
the photometric errors in the process of identifying the optimal period. 

Saha et al's (1990) paper appears to be the only previous study of the 
RR Lyrae population in NGC 147. We are motivated to revisit the 
properties of these stars for two reasons. First, we have had good 
success in using a light curve template-fitting
algorithm (Layden \& Sarajedini 2000) to determine the properties of 
RR Lyraes such as periods, amplitudes, and mean magnitudes. This method 
includes the photometric errors in the analysis and has been
streamlined and redesigned to be more user-friendly by 
Mancone \& Sarajedini (2008). This will allow us to refine the determination 
of the RRL
periods and place better constraints on the total number of such variables 
in NGC 147. Second, there are archival imaging data from the Hubble
Space Telescope (HST) Wide Field Planetary Camera 2 (WFPC2) for NGC 147 
that may allow us to update the list of RR Lyraes published by S90. 
Although the WFPC2 data provide accurate photometry at the level of 
the HB in NGC 147, they exhibit poor phase coverage with a time baseline
of $\sim$0.4 day. However, it is still useful because it could help to identify 
the shorter period, lower amplitude RR Lyrae stars that may not have been 
detected by S90.

The next section describes the observational datasets that we will analyze.
Sections 2 and 3 make it clear that neither the S90 data nor the WFPC2 
data are  ideal for the purpose of studying the RR Lyraes
in NGC 147, but they do complement each other nicely. As discussed in 
Section 4, it is important to carry out simulations to fully understand the 
biases and caveats inherent in the results obtained from each of the 
datasets. The conclusions are presented in Section 5 as well as the 
case for future work to better characeterize
the variable stars in NGC 147 and its M31 dwarf satellite cousin NGC 185.



\section{Observations and Data Reduction}

\subsection{Ground Based Data}

The $g$-band (Thuan \& Gunn 1976) photometry of NGC~147 is available in 
data tables provided by S90. The target field, located $6'$ northwest 
of the galaxy's center along the semi-major axis, was observed 
15 times with 20 min exposures in 1986, and 8 times with 30 min exposures in
1987, both using the Hale 5m telescope with the 4-shooter CCD. The 30 min 
exposures were obtained during four consecutive nights. The limiting magnitude 
of their photometry is $g=26.5$, which is about 1.25 magnitude fainter than the 
HB magnitude of NGC~147. The photometric completeness of the HB 
stars (at $g$$\sim$25.5) is 62$\%$ in the crowded regions and 72$\%$ in 
those that are less crowded.  The photometric errors for the HB stars are
$\sigma > 0.15$ mag, which is comparable to the amplitude of c-type 
or low amplitude ab-type RR Lyrae variables. 
Bailey ab-type stars are fundamental mode pulsators which have a
sawtooth-like, asymmetrical light curve shape, while first-overtone c-type
pulsators have sinusoidal light curves.  RRL with ab-type (RRab) have
longer periods (0.5 $<$ P $<$ 1.2 days) than the c-types (RRc) stars 
(0.2 $<$ P $<$ 0.5 days; Smith 1995)

\subsection{Hubble Space Telescope Data}

Observations taken with HST/WFPC2 of the outer regions of NGC 147 
are available in the HST archive (Han et al, 1997, program ID : GO-6233). The 
target field, located at 4$'$ south of the galaxy's center, was imaged 7 times 
in F555W ($\sim$V) and 6 times in F814W ($\sim$I), with exposure times from 
1300s to 2800s as detailed in Table 2. The time baseline of the dataset spans 
0.4 days. The observations 
were primarily intended to study the characteristics of the (nonvariable)
stellar populations of NGC 147. 

All of the WFPC2 images were photometered by using the HSTphot package 
(Dolphin 2000), which is designed for use on WFPC2 data. First, any 
image defects, such as bad pixels, cosmic rays, and hot pixels were removed by 
using the utility software included within HSTphot. Then, the photometric 
measurements were made on each image by running HSTphot in ``PSF fitting"
mode. HSTphot uses a library of point spread functions (PSFs) for WFPC2 images. 
The aperture corrections, defined by the average difference between aperture 
photometry with a 0.5" radius and the PSF photometry, were also applied by HSTphot. 
A minimum threshold for object detection was set 3$\sigma$ above the background 
signal. We selected stars with high-quality photometry (i.e. stellar profiles with
$\sigma_{V,I} < 0.15$ mag) from the output of HSTphot 
for further analysis. 

Figure 1 shows the VI CMD of NGC~147 from the HST/WFPC2 data. The photometric 
limit reaches V $\sim$ 28.2 mag. The broad red giant branch (RGB) of NGC 147 
primarily indicates a wide range of metallicity among the stellar populations in this 
galaxy. This figure also shows a blue horizontal branch (BHB) and a distinct instability 
strip gap which is the well known location of RR Lyrae variable stars. We performed 
completness tests of our photometry using HSTphot's artificial star feature. This module
creates 
a comparable number of artificial stars for each color-magnitude bin from the observed 
CMD and randomly distributes these in each of the original WFPC2 image. The artificial 
stars are photometered in exactly the same manner as the actual stars. Figure 2 
illustrates the result of these completeness tests. At the magnitude of the HB 
(V $\sim$ 25.5 mag), the completeness level is about 95\%. Therefore, we do
not expect photometric incompletness to adversely affect the detection of RR Lyrae
candidates identified in the WFPC2 data. 

\section{RR Lyrae Period Determination}  

The previous study (Saha et al 1990) used the Lafler-Kinman (L-K) algorithm 
(Lafler \& Kinman 1965), a prototype of the phase dispersion minimization method 
(PDM) for period determination. The L-K algorithm defines a test parameter, $\Theta$, 
the sum of the squares of the differences between two adjacent magnitudes rearranged 
in the ascending order of phases for each trial period. The algorithm then searches for 
the period that minimize the test parameter. Since the L-K method uses only one free 
parameter (period) for the period optimization, the calculation is relatively simple, 
straightforward, and faster than other period finding routines. However, it does not take 
into account some important factors, such as the amount of photometric error. Large
photometric errors can effectively mask intrinsic variability present in the data
making subsequent period solutions highly suspect.
As a result, as Lafler \& Kinman stressed in their original paper, the L-K algorithm may have 
difficulty in obtaining reliable periods for variables with relatively small amplitudes 
($\lea$ 0.75 mag) when the data set has a limited number of observations and moderate 
photometric accuracy. Because the majority of RR Lyrae variables have amplitudes
that fall below this threshold, it is important to be cautious in interpreting periods and 
amplitudes derived from the L-K method especially when the photometric errors
are significant. 

We have analyzed both the HST WFPC2 and the $g$-band photometric datasets 
(Saha et al. 1990) using our own light curve template-fitting software dubbed 
`FITLC' (Layden 1998; Mancone \& Sarajedini 2008). It uses an algorithm 
known as `Pikaia' (Charbonneau 1995), which is a robust optimization 
routine that computes the best combination of period and amplitude in order to 
minimize the $\chi^2$ values between the observed data points and 10 different 
light curve templates taken from the work of Layden (1998). 
  
The advantages of using template fitting for the determination of RR Lyrae periods 
are twofold. First, based on extensive tests we have performed, the template-fitting 
routine generally works better for small numbers (N $\lea$ 30) of observations than 
the L-K method. The power of the template-fitting method for the determination of 
light curve properties from relatively small numbers of points is well presented in the
literature (Mackey \& Gilmore 2003; Sarajedini et al 2006; Mancone \& 
Sarajedini 2008). 

 
Secondly, the template
fitting method provides an assessment of the variable star classification based on
the shape of the phased light curve.  Whether the RRL is an ab-type
or c-type naturally follows from the results
of the method based on which template provides the best fit to the
observational data.

\subsection{Reanalysis of the Saha et al. Data}

Given the advantages of template light curve fitting over the L-K 
algorithm for cases where the number of observations is small, we have 
reanalyzed the $g$-band photometry of 32 RR Lyrae candidates in NGC 147 
from S90 using the FITLC routine. 
  Figures 3 through 11 show the best fitting light curves for the RR 
Lyrae candidates from FITLC as compared with the results from S90. 
The actual periods determined by the two methods are compared in 
Fig. 12 and listed in Table 3, which also shows the classification of
each variable star and the mean magnitude.
Overall, the resulting periods from the L-K and 
FITLC methods applied to the S90 $g$-band data bear little resemblance to each other.
We found only 6 cases (C1-V9, C1-V10, C1-V11,C3-V8, C3-V12, and C4-V9) out 
of 32 RR Lyraes where L-K and FITLC agree reasonably well in both period and 
classification. However, we found that L-K periods tend to be longer than FITLC 
periods, especially in the short period (P $<$ 0.6 days) and lower
amplitude ranges (Amplitude $\lea$0.75). 
  
In order to further investigate the differences we see between the L-K method 
and FITLC, we have performed the following set of simulations. We generated
eighth-order Fourier decompositions of the template light curves of Layden (1998), 
6 ab-types and 2 c-types, and calculated their Fourier parameters. With the
functional forms of these 8 light curves, 
we created synthetic RR Lyrae light 
curves with known periods and amplitudes that mimic the $g$-band observations
of S90. We sampled each light curve at 23 different epochs (the maximum
number in the S90 analysis) with measurement
errors given by the mean value taken from the actual $g$-band data.  
Periods and amplitudes were randomly assigned to each synthetic RR Lyrae 
from reasonable period-amplitude ranges for
typical RR Lyrae variables 
( 0.2 $<$ P $<$ 1.2 days, 0.2 $<$ Amp $<$ 1.5 mag). In this way, 
the artificial RR Lyrae variables properly represent important observational 
conditions that can affect the period determination such as the time baseline, 
the number of observations, and the photometric errors. We then applied the
L-K and FITLC methods to the synthetic light curves in exactly the same
manner as our original period finding routines and compared the derived
periods with the input ones. 
 
Figures 13 (L-K method) and 14 (FITLC) illustrate the results of our 
simulations for the $g$-band observations of S90. 
They  show that FITLC works better than the L-K method in 
finding periods using the S90 photometry;
however some systematic errors still remain in the FITLC results. 
The FITLC routine recovered $\sim$59\% of the input periods from the 
synthetic RR Lyraes with a period error of $\pm$ 0.1 days, 
while the L-K method only recovered $\sim$15\% of the input periods 
with the same period error. 
Indeed, even though the $g$-band photometry of NGC147 from S90 covered a significant
observational baseline ($\sim$4 to 5 days) 
and provided a reasonable number of observations ($\leq$23 epochs), 
the period finding results are largely unreliable because of the relatively
large errors in the photometry. 
If we reduce the errors by a factor of 2, then the recovery efficiency
of the input periods from the FITLC routine 
is enhanced up to $\sim$72\% but remains unchanged for the L-K method. 
We should note that the average photometric error of the $g$-band photometry at the
level of the HB is $\sim$0.2 mag. 
Even if we reduce the magnitude error by a factor of 2, most of the photometric data
still exhibit errors of $\sim$0.1 mag,
which is still too large to facilitate accurate period determination. 
Thus, the properties of the RR Lyrae candidates derived from the
$g$-band photometry must be interpreted with 
extreme caution.

\subsection{Analysis of HST/WFPC2 Archival Data} 

As shown in section 2.2, the HST WFPC2 data of NGC 147 are deep enough 
to provide accurate V magnitudes for RR Lyrae candidates. The average photometric 
error in the V band at the HB magnitude level (V$\sim$25.5) is $<$0.05 mag. 
Therefore, unlike the $g$-band photometry of S90, period determination from the
WFPC2 data should not be adversely affected by the photometric errors. 
However, the WFPC2 archival data for NGC 147 have a short observational 
baseline ($\sim$0.4 days) with a small number of available epochs. The c-type 
RR Lyraes, which have periods of $\sim$0.2 - 0.4 days with relatively small 
amplitudes ($\lea$ 0.3 mag ), might be less affected by this short observational 
baseline. 

Given these limitations of the data, we proceeded with caution in defining
our set of candidate RR Lyraes. First, we selected stars with a color-magnitude 
range ($-1 < (V-I) <1$, and $24.5 < V < 26$), shown as a rectangular box in 
Figure 1. Then, we calculated the reduced $\chi^2$ of the observed 
V and I magnitudes of each star as a variability index defined by the 
following formula

\begin{displaymath}
\chi^2 = \frac{1}{N_V + N_I} \times
\Bigg[\sum_{i=1}^{N_V} \frac{(V_i - \overline V)^2}{\sigma_i^2} +  
 \sum_{i=1}^{N_I} \frac{(I_i - \overline I)^2}{\sigma_i^2}\Bigg].
\end{displaymath}

\noindent This diagnostic is distinct from the variability index of Welch \&
Stetson (1993) because the latter uses correlations in variability between
different filter passbands. 
Stars with $\chi^2$ values greater than 2.0 were considered as variable candidates. 
Based on this criterion, 931 stars were selected. We applied the FITLC template 
light curve fitting routine to the $V$- and $I$-band observations of these variable 
candidates in order to find the best combination of period and amplitude.  
Of these, 36 RR Lyraes
(32 ab-type and 4 c-type) have colors and magnitudes that place them along the 
HB where we would expect RR Lyraes to be located.
Figure 15, 16, and 17  show the best fitting light curves for these RR Lyrae 
candidates and Fig. 18 illustrates their location in the NGC 147 CMD. 
Table 4 gives their positions as measured from the world coordinate system
information in the image headers as well as their individual mean magnitudes and
colors.
%
The mean V magnitude of all of the RR Lyrae candidates is 
$<V> = 25.40 \pm 0.16$. Given
the mean metallicity derived by Nowotny et al. (2003) and the RR Lyrae luminosity-metallicity
relation from Chaboyer (1999), $M(RR) = 0.23[Fe/H] + 0.93$, we find a 
distance modulus of $(m-M)_0 = 24.16 \pm 0.16$ by applying a reddening of
E(B--V) = 0.18 from Schlegel et al (1998). This value is in reasonable agreement with
the distance quoted in Table 1 from Sharina et al. (2006) and the value of
$(m-M)_0 = 24.39 \pm 0.05$ from Han et al. (1997); our distance places NGC 147
approximately 100 kpc in front of M31. 

In order to assess the effects of the observing window on the 
FITLC periods derived from the WFPC2 photometry, we carried out synthetic 
light curve simulations as we did on the ground-based data of Saha et al. (1990). 
Hundreds of artificial RR Lyraes were created using template light 
curves with known periods and amplitudes. The number of epochs, magnitudes 
and photometric errors were carefully assigned to mimic the
WFPC2 data. Then, we applied the FITLC algorithm to this simulated dataset. 
The simulation results are presented in Figure 19. They show that the WFPC2 
time-series data of NGC 147 are better suited for the
investigation of RR Lyrae periods than the ground-based $g$-band photometry
from Saha et al. (1990). However, that being said, for the entire 
range of test periods (0.2 d $<$ P $<$ 1.2 d), only $\sim$39 \% of the output periods 
are in agreement with the input periods of the artificial RR Lyraes within 
$\pm$0.1 days. The situation is markedly improved for the short period variables 
(P $<$ 0.40 d), which tend to be the c-type RR Lyraes. They exhibit overall
better period recovery ($\sim$ 84 \%) within the given period errors.

Generally speaking however, our investigation of both the $g-$band 
ground-based photometry and the VI HST/WFPC2 
archival data leads us to the conclusion that our current knowledge of the periods and
amplitudes of RR Lyrae stars in NGC 147 is highly uncertain. We need better 
observational data with high accuracy photometry, sufficient coverage of the 
observational baseline, and many available epochs, in order to fully understand 
the characteristics of the RR Lyrae population of NGC 147.

\section{Summary and Conclusions}

In this study, we present an investigation of the RR Lyrae population
in the local group dwarf galaxy, NGC 147 using available time-series photometry 
from both the ground and HST. Based on our period finding analysis and artificial 
variable star tests, we draw the following conclusions:

1. The $g$-band photometry from the work of Saha et al. (1990) likely possesses
an adequate observational baseline and available epochs. However, our simulations 
showed that the photometric errors at the level of the horizontal branch
significantly hinder the accurate determination of the pulsation periods of 
the RR Lyrae candidates in NGC 147. 

2. Our template light curve fitting technique (FITLC) detected 36 probable 
RR Lyrae candidates from HST/WFPC2 archival data. However, our 
simulations reveal that the short observational baseline and small number 
of observations severely affect the accurate characterization of 
RR Lyrae periods longer than $\sim$0.4 days, which are essentially the
ab-type RR Lyraes.

3. The $g$-band photometry and the WFPC2 archival data analyzed herein
present two extreme cases often found in period finding studies - good phase 
coverage but with large photometric errors, and high quality photometry with 
poor phase coverage. Our investigation of these two extreme cases not 
only provides a good reference for interpreting the pulsation properties of 
RR Lyrae variables in other similar situations, but also calls attention to a 
strong need for new high quality time-series observations of NGC 147.
Thus, while we can confidently assert that NGC 147 contains RR Lyrae
variables, and therefore a population older than $\sim$10 Gyr, it is
not possible at this time to use the pulsation properties of these RR Lyraes
to study other aspects of this old population.
 
\acknowledgments

We thank Karen Kinemuchi for detailed comments on an early version
of this manuscript.
We gratefully acknowledge support from NASA through grant 
AR-11277.01-A from the Space Telescope Science Institute, which is 
operated by the Association of Universities
for Research in Astronomy, Inc., for NASA under contract NAS5-26555.






\clearpage

\begin{deluxetable}{lcc}
\tablecaption{Physical properties of NGC 147.\label{tbl-1}} 
\tablewidth{0pt}
\tablehead{
\colhead{Property}
  &\colhead{Value}
  &\colhead{Reference}
}
\startdata
RA           & 00$^h$ 33$^m$ 11.9$^s$      &   \\
Dec          &  +48$^{\circ}$ 30$'$ 24.8$"$      &   \\
($l$, $b$)   & 119.82$^{\circ}$, -14.25$^{\circ}$    &   \\
V$_r$        & -193$\pm$3 $km/s$              &  Bender et al 1991\\
(m-M)$_0$    & 23.95                                   &  Sharina et al 2006\\
M$_{V,0}$    &  -15.1                                    & "  \\
E(B-V)       & 0.18$\pm$0.03                       &  Schlegel et al 1998\\
A$_V$        & 0.580                                      & " \\
$<[Fe/H]>_{RGB}$  & -1.11$\pm$0.01     &  Nowotny el al 2003 \\
$<[Fe/H]>_{GCs}$  & -2.2$\pm$0.42        & Da Costa \& Mould 1988 \\
\enddata
\end{deluxetable}

\begin{deluxetable}{lccc}
\tablecaption{WFPC2 Observing Log\label{tbl-2}} 
\tablewidth{0pt}
\tablehead{
\colhead{Dataset}
  &\colhead{Filter}
  &\colhead{Exp time}
  &\colhead{2 450 000 -- HJD}
}
\startdata
u2ob0101t   &  F555W & 2400s & 9890.01465 \\
u2ob0102t   &  F555W & 1300s & 9890.06934 \\
u2ob0103t   &  F555W & 1300s & 9890.08594 \\
u2ob0104t   &  F555W & 1300s & 9890.13672 \\
u2ob0105t   &  F555W & 1300s & 9890.15332 \\
u2ob0106t   &  F555W & 2800s & 9890.21289 \\
u2ob0107t   &  F555W & 2800s & 9890.27930 \\
u2ob0108t   &  F814W & 1300s & 9890.33789 \\
u2ob0109t   &  F814W & 1300s & 9890.35449 \\
u2ob010at   &  F814W & 1300s & 9890.40527 \\
u2ob010bt   &  F814W & 1300s & 9890.42188 \\
u2ob010ct   &  F814W & 1300s & 9890.47168 \\
u2ob010dt   &  F814W & 1300s & 9890.48828 \\
\enddata
\end{deluxetable}

\clearpage

\begin{deluxetable}{lcccccc}
\tablecaption{Characteristics of the Saha et al. (1990) Variables\label{tbl-a}}
\tablewidth{0pt}
\tablehead{
\colhead{Object}
  &\colhead{P(LK)}
  &\colhead{$<g>$(LK)}
  &\colhead{Type(LK)\tablenotemark{1}}
  &\colhead{P(FITLC)}
  &\colhead{$<g>$(FITLC)}
  &\colhead{Type(FITLC)\tablenotemark{1}}
}
\startdata
 C1-V1  & 0.58543  & 24.65 & ab & 0.21990 & 24.60 & EB \\
 C1-V2  & 0.49480  & 24.71 & ab & 0.98318 & 24.69 & EB \\
 C1-V3  & 0.42978  & 25.05 & ab & 0.28723 & 24.95 & EB \\
 C1-V4  & 0.72259  & 25.06 & ab & 0.30274 & 24.86 & c \\
 C1-V5  & 0.38861  & 24.62 & c? & 0.37888 & 25.20 & ab \\
 C1-V6  & 0.28260  & 24.46 & c? & 0.64538 & 24.20 & c \\
 C1-V7  & 0.34835  & 24.55 & c? & 0.41998 & 24.58 & c \\
 C1-V8  & 0.81660  & 24.95 & ab & 0.84910 & 24.88 & c \\
 C1-V9  & 0.53655  & 25.04 & ab & 0.53753 & 25.08 & ab \\
 C1-V10 & 0.86053  & 24.68 & ab & 0.85044 & 24.86 & ab \\
 C1-V11 & 0.43104  & 24.80 & ab & 0.43352 & 24.79 & ab \\
 C1-V12 & 0.27895  & 24.81 & c  & 0.55790 & 24.91 & EB \\
 C1-V13 & 0.71724  & 24.99 & ab & 0.41037 & 24.95 & ab \\
 C1-V14 & 0.27355  & 25.11 & c  & 0.21888 & 25.04 & EB \\
 C3-V1  & 0.52933  & 25.44 & ab & 0.51879 & 25.21 & c \\
 C3-V2  & 0.86729  & 25.25 & ab & 0.85738 & 25.18 & c \\
 C3-V3  & 0.74508  & 24.54 & ab & 0.68766 & 24.56 & ab \\
 C3-V4  & 0.60875  & 24.77 & ?  & 0.30546 & 24.80 & c \\
 C3-V5  & 0.54649  & 25.18 & ab & 0.36870 & 25.37 & ab \\
 C3-V6  & 1.22297  & 24.84 & AC & 0.76788 & 24.88 & ab \\
 C3-V7  & 1.23533  & 24.53 & AC & 0.55284 & 24.57 & ab \\
 C3-V8  & 0.54194  & 24.54 & ab & 0.53806 & 24.53 & ab \\
 C3-V9  & 0.69967  & 25.01 & ab & 0.34991 & 24.88 & c \\
 C3-V10 & 0.67132  & 25.17 & ab & 0.33587 & 25.36 & ab \\
 C3-V11 & 0.57346  & 24.69 & ab & 0.45515 & 24.70 & ab \\
 C3-V12 & 0.71666  & 25.00 & ab & 0.75819 & 24.96 & ab \\
 C3-V13 & 0.75816  & 24.70 & EB? & 0.70529 & 24.65 & c \\
 C4-V1  & 0.77865  & 24.76 & ab & 0.57333 & 24.68 & ab \\
 C4-V2  & 0.75304  & 25.58 & ab & 0.59890 & 25.38 & c \\
 C4-V3  & 0.76366  & 25.38 & ab & 0.40267 & 25.59 & ab \\
 C4-V4  & 0.46348  & 24.96 & ab & 0.82539 & 25.02 & c \\
 C4-V5  & 0.77979  & 25.28 & ab & 0.60954 & 25.02 & c \\
 C4-V6  & 0.64373  & 25.21 & ab & 0.67994 & 25.30 & c \\
 C4-V7  & 0.57137  & 25.43 & ab & 0.34560 & 25.52 & ab \\
 C4-V8  & 0.29731  & 25.22 & c  & 0.36375 & 25.08 & EB \\
 C4-V9  & 0.60459  & 25.35 & ab & 0.55696 & 25.31 & ab \\
\enddata

\tablenotetext{1}{ab = ab-type RR Lyrae, c = c-type RR Lyrae,
EB = eclipsing binary, AC = anomalous cepheid}

\end{deluxetable}

\clearpage

\begin{deluxetable}{lcccc}
\tablecaption{Characteristics of RR Lyrae Candidates\label{tbl-b}}
\tablewidth{0pt}
\tablehead{
\colhead{Object}
  &\colhead{RA(J2000.0)}
  &\colhead{Dec(J2000.0)}
  &\colhead{$<V>$}
  &\colhead{$<V-I>$}
}
\startdata
 V18120 &   0 33 13.41 & 48 28 50.22 &  25.137  &   0.537 \\  
 V15719 &   0 33 11.18 & 48 28 06.05 &  25.072  &   0.753 \\ 
 V16803 &   0 33 12.61 & 48 27 51.49 &  25.079  &   0.740 \\ 
 V17600 &   0 33 13.28 & 48 28 23.10 &  25.224  &   0.529 \\  
 V18304 &   0 33 09.60 & 48 28 10.86 &  25.169  &   0.331 \\ 
 V20325 &   0 33 09.83 & 48 28 11.68 &  25.293  &   0.278 \\
 V20731 &   0 33 09.20 & 48 28 21.55 &  25.459  &   0.435 \\ 
 V20834 &   0 33 11.69 & 48 28 50.55 &  25.309  &   0.588 \\ 
 V22046 &   0 33 14.32 & 48 28 27.68 &  25.389  &   0.447 \\ 
 V22251 &   0 33 13.13 & 48 27 45.58 &  25.431  &   0.204 \\ 
 V22308 &   0 33 13.43 & 48 27 44.47 &  25.484  &   0.099 \\ 
 V22660 &   0 33 13.26 & 48 28 37.32 &  25.392  &   0.155 \\ 
 V23104 &   0 33 11.66 & 48 28 47.19 &  25.632  &   0.682 \\ 
 V23505 &   0 33 14.28 & 48 28 22.03 &  25.588  &   0.519 \\ 
 V24080 &   0 33 12.82 & 48 28 56.96 &  25.657  &   0.219 \\ 
 V37537 &   0 33 14.89 & 48 27 32.23 &  25.161  &   0.636 \\ 
 V38170 &   0 33 16.90 & 48 27 08.75 &  25.234  &   0.684 \\
 V38302 &   0 33 19.96 & 48 27 53.65 &  25.300  &   0.727 \\
 V38757 &   0 33 22.22 & 48 27 43.15 &  25.317  &   0.170 \\
 V38944 &   0 33 15.98 & 48 27 20.24 &  25.303  &   0.538 \\
 V39451 &   0 33 23.75 & 48 27 40.43 &  25.338  &   0.508 \\
 V39591 &   0 33 17.03 & 48 28 01.65 &  25.368  &   0.596 \\
 V39668 &   0 33 19.29 & 48 27 46.54 &  25.440  &   0.792 \\  
 V39906 &   0 33 21.73 & 48 27 35.56 &  25.435  &   0.420 \\ 
 V40264 &   0 33 20.47 & 48 27 52.46 &  25.471  &   0.600 \\  
 V40315 &   0 33 20.74 & 48 27 36.23 &  25.510  &   0.596 \\  
 V40378 &   0 33 18.59 & 48 27 36.31 &  25.513  &   0.418 \\  
 V41094 &   0 33 18.34 & 48 27 10.21 &  25.568  &   0.412 \\ 
 V41232 &   0 33 18.79 & 48 27 21.51 &  25.600  &   0.434 \\
 V56535 &   0 33 14.15 & 48 26 39.45 &  25.395  &   0.516 \\
 V56641 &   0 33 11.18 & 48 26 17.77 &  25.406  &   0.409 \\
 V56868 &   0 33 14.99 & 48 25 42.19 &  25.399  &   0.329 \\
 V57185 &   0 33 13.88 & 48 26 36.28 &  25.519  &   0.714 \\
 V57231 &   0 33 14.95 & 48 25 51.15 &  25.498  &   0.650 \\ 
 V57831 &   0 33 16.45 & 48 25 56.11 &  25.684  &   0.721 \\ 
 V58117 &   0 33 14.23 & 48 26 14.76 &  25.639  &   0.310 \\  
\enddata
\end{deluxetable}

\begin{figure}
\epsscale{1.0}
\caption{VI color-magnitude diagram for the dwarf elliptical galaxy, NGC 147. 
A color-magnitude range ($-1 < (V-I) <1$, and $24.5 < V < 26$) shown as a 
box is used to search for RR Lyrae candidates. \label{fig1}}
\end{figure}

\clearpage

\begin{figure}
\epsscale{1.0}
\plotone{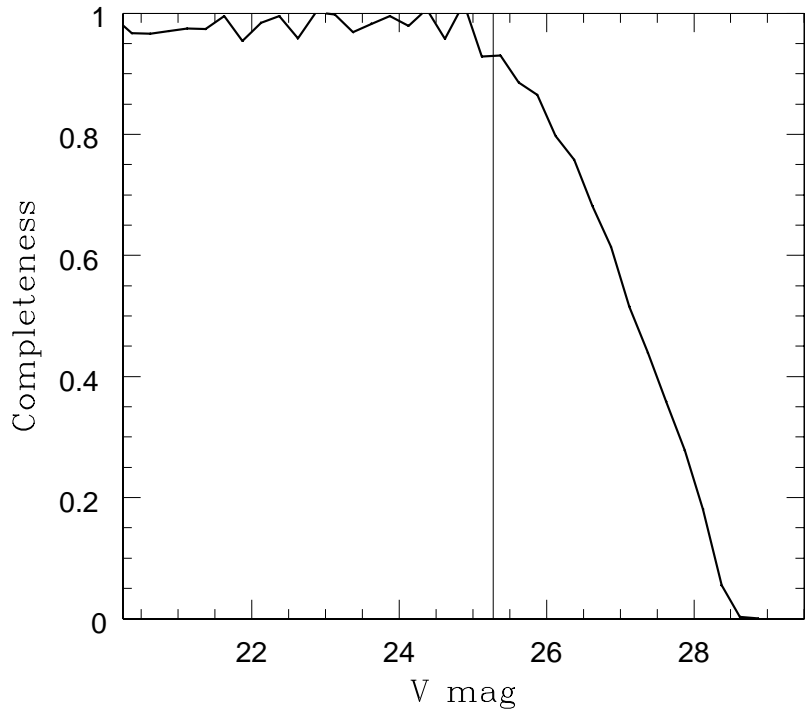}
\caption{Completeness test for the WFPC2 VI photometry. A vertical line indicates the horizontal branch (HB) magnitude of NGC 147. The photometric completeness of HB stars reaches almost 93$\%$.  \label{fig2}}
\end{figure}

\clearpage

\begin{figure}
\epsscale{1.0}
\plotone{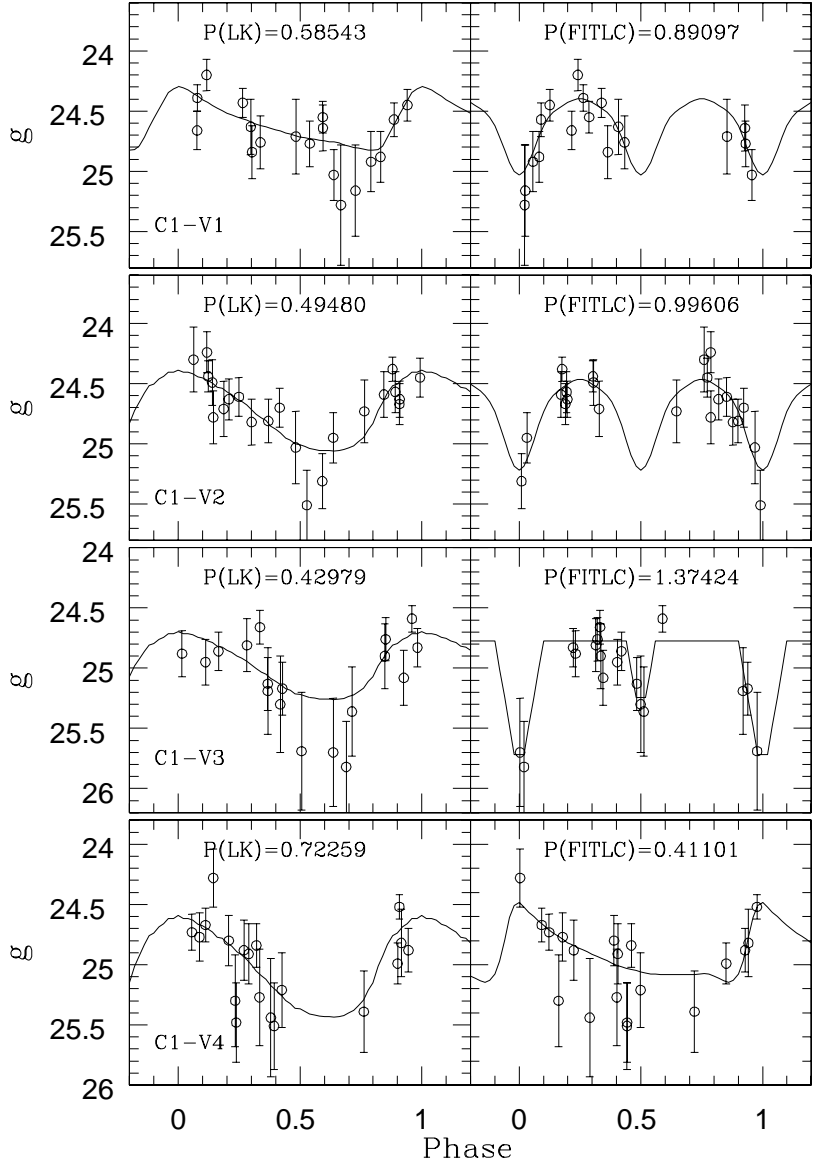}
\caption{ The left panels show the RR Lyrae light curves phased using the
period derived by Saha et al. (1990) via the L-K period determination
method. The right panels show the same variables but phased using 
periods derived from the templated-fitting FITLC method. In this
and subsequent figures, it is apparent that a number of stars thought
to be RR Lyrae variables could in fact be eclipsing or contact
binaries. \label{fig3}}
\end{figure}

\clearpage

\begin{figure}
\epsscale{1.0}
\plotone{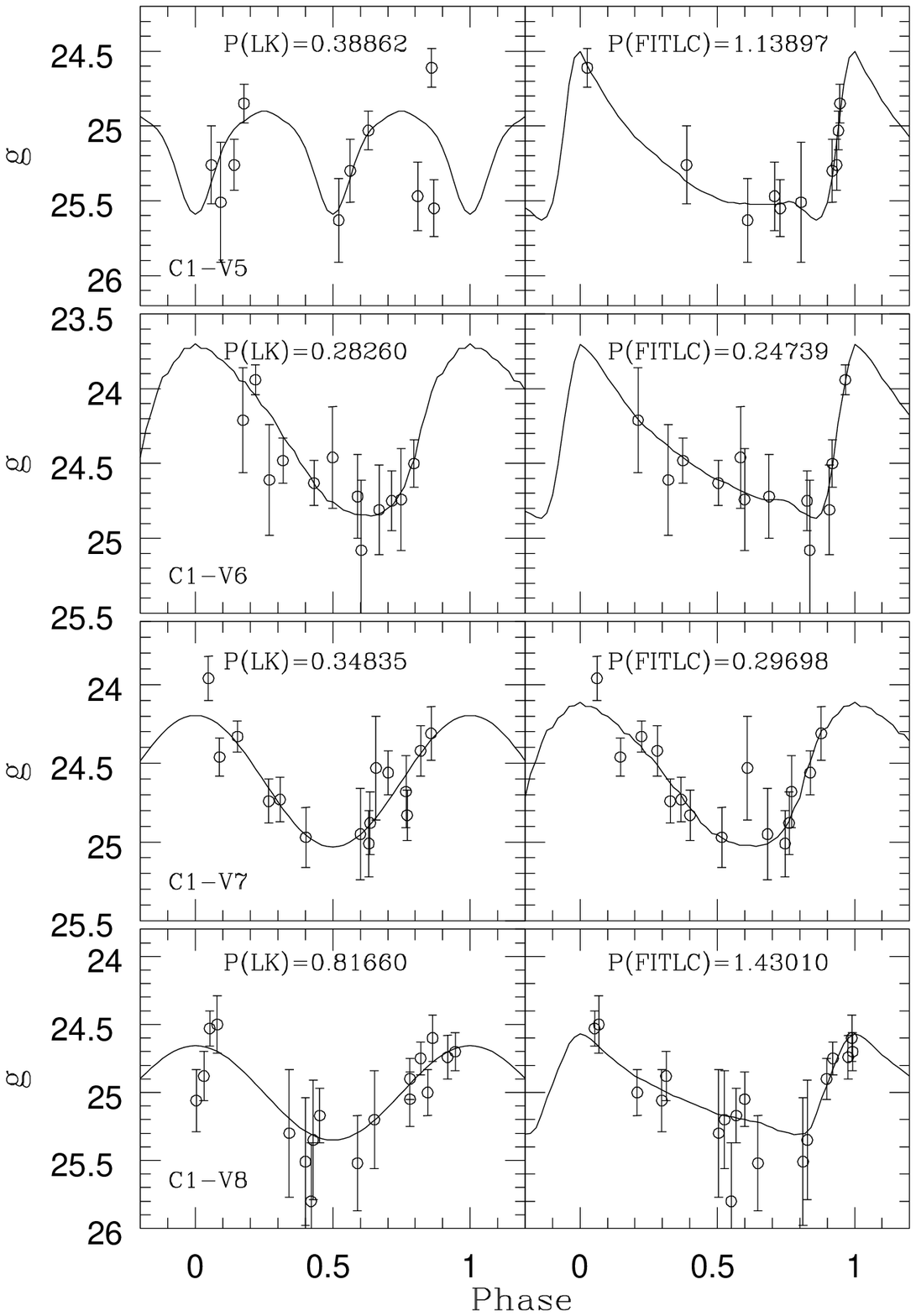}
\caption{same as Figure 3. \label{fig4)}}
\end{figure}

\clearpage

\begin{figure}
\epsscale{1.0}
\plotone{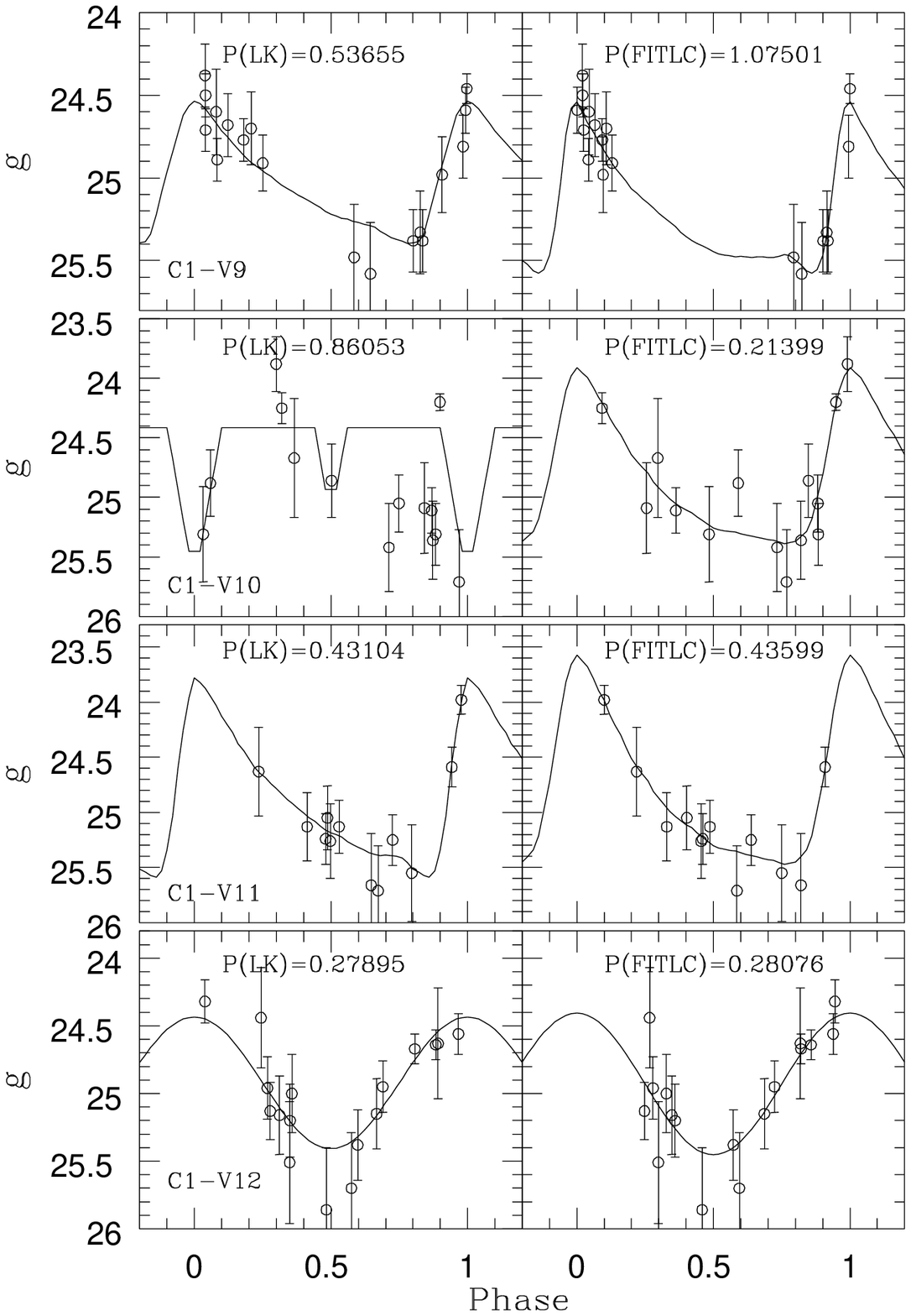}
\caption{same as Figure 3. \label{fig5)}}
\end{figure}

\clearpage

\begin{figure}
\epsscale{1.0}
\plotone{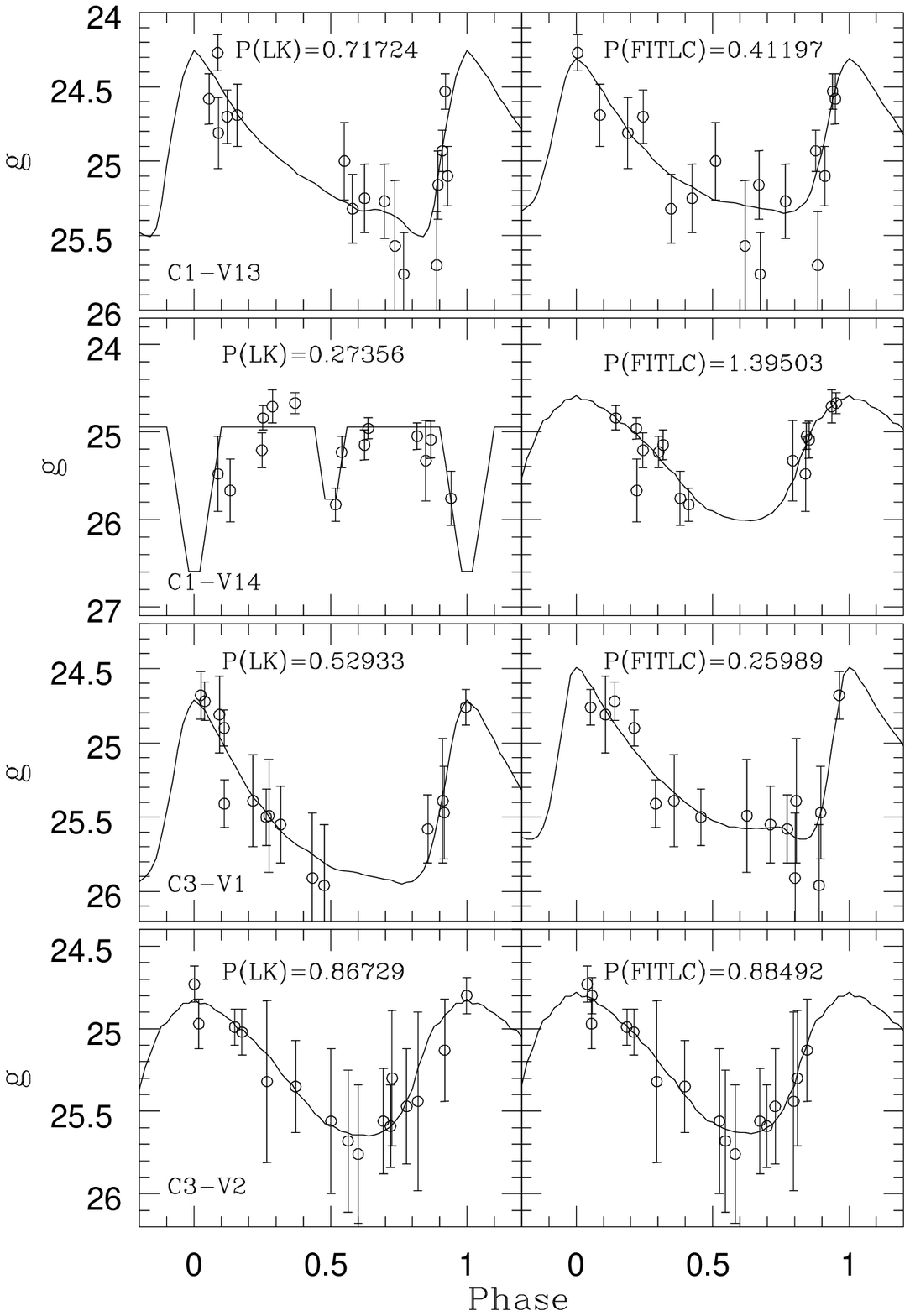}
\caption{same as Figure 3. \label{fig6)}}
\end{figure}

\clearpage

\begin{figure}
\epsscale{1.0}
\plotone{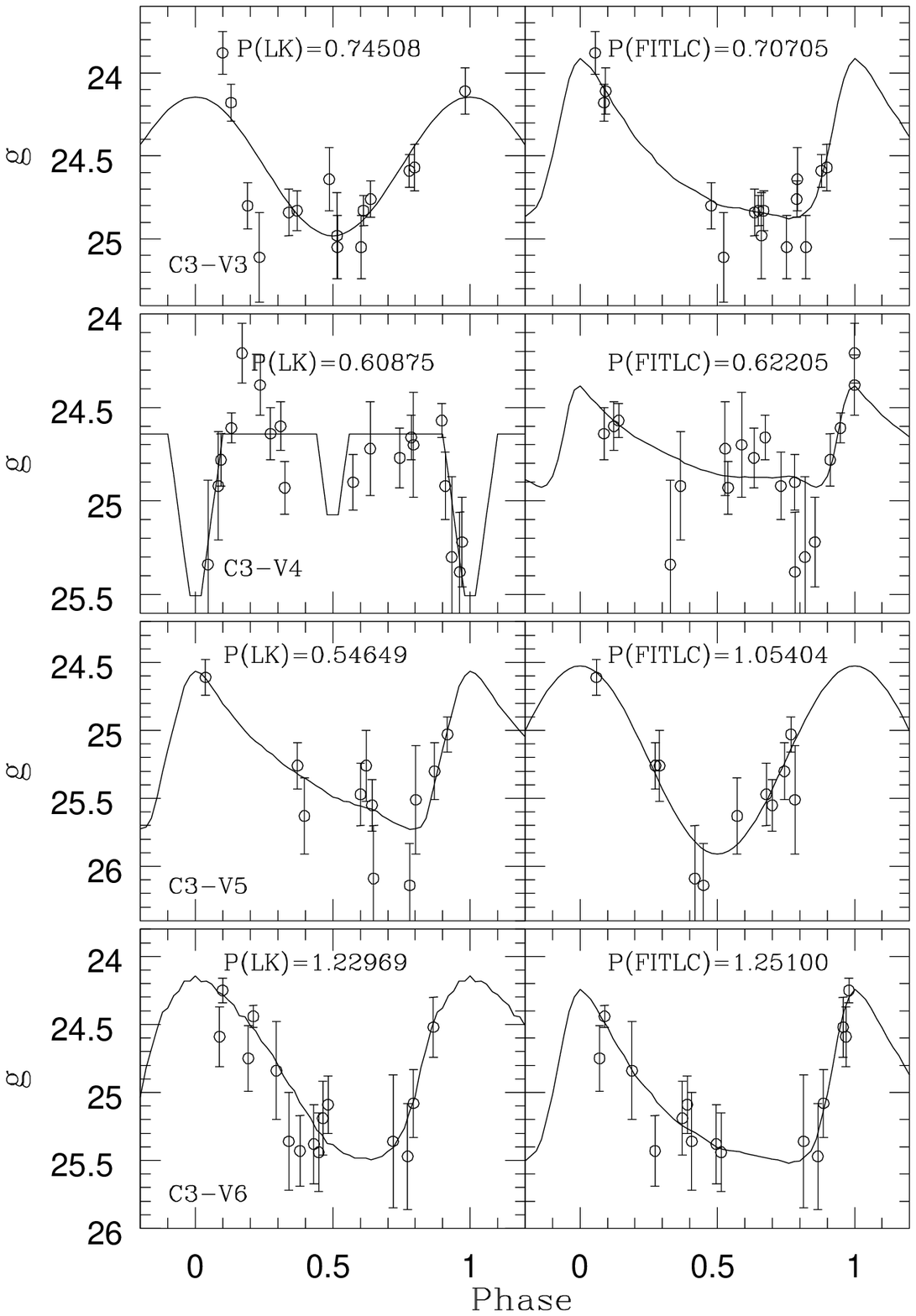}
\caption{same as Figure 3. \label{fig7)}}
\end{figure}

\clearpage

\begin{figure}
\epsscale{1.0}
\plotone{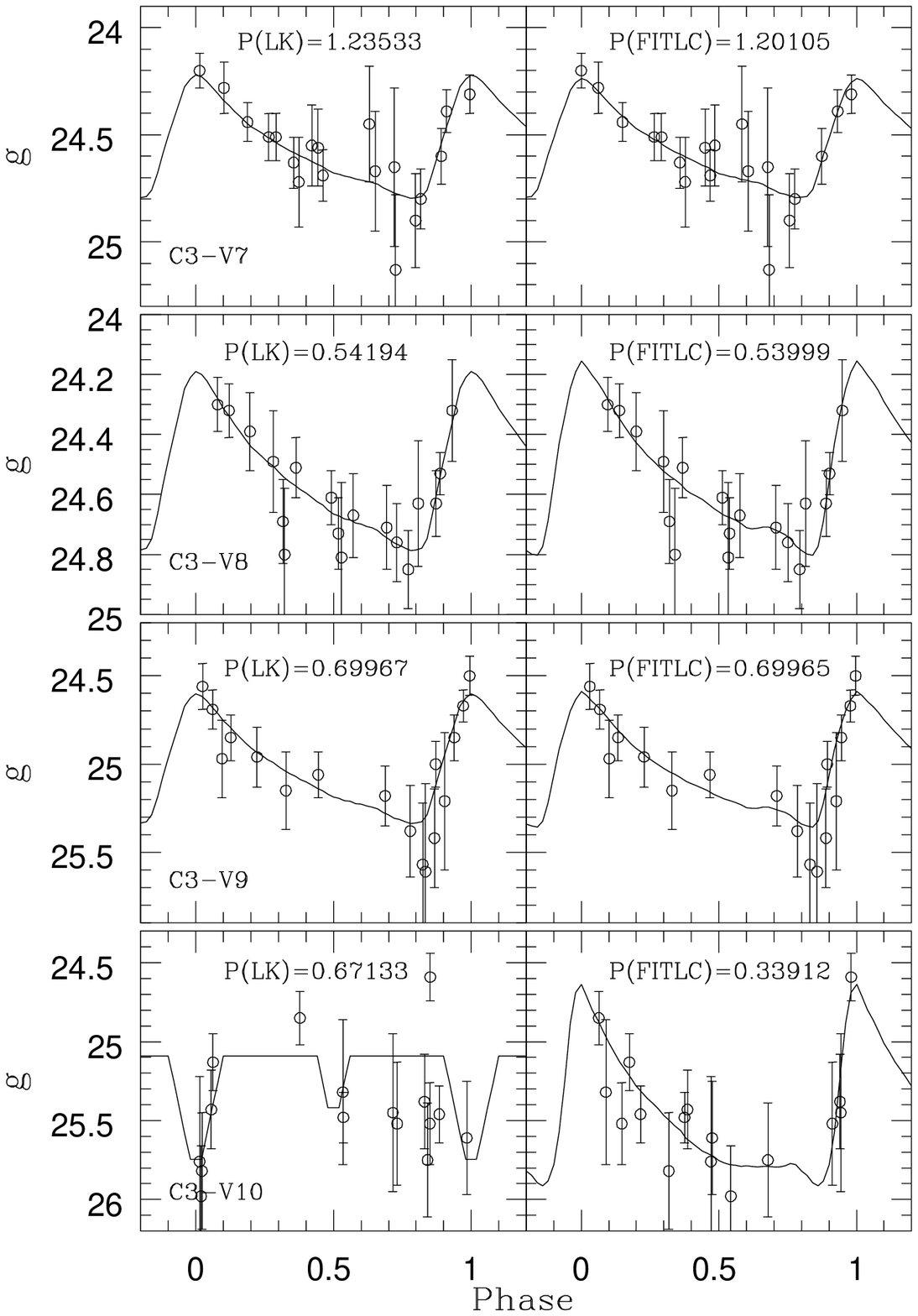}
\caption{same as Figure 3. \label{fig8)}}
\end{figure}

\clearpage

\begin{figure}
\epsscale{1.0}
\plotone{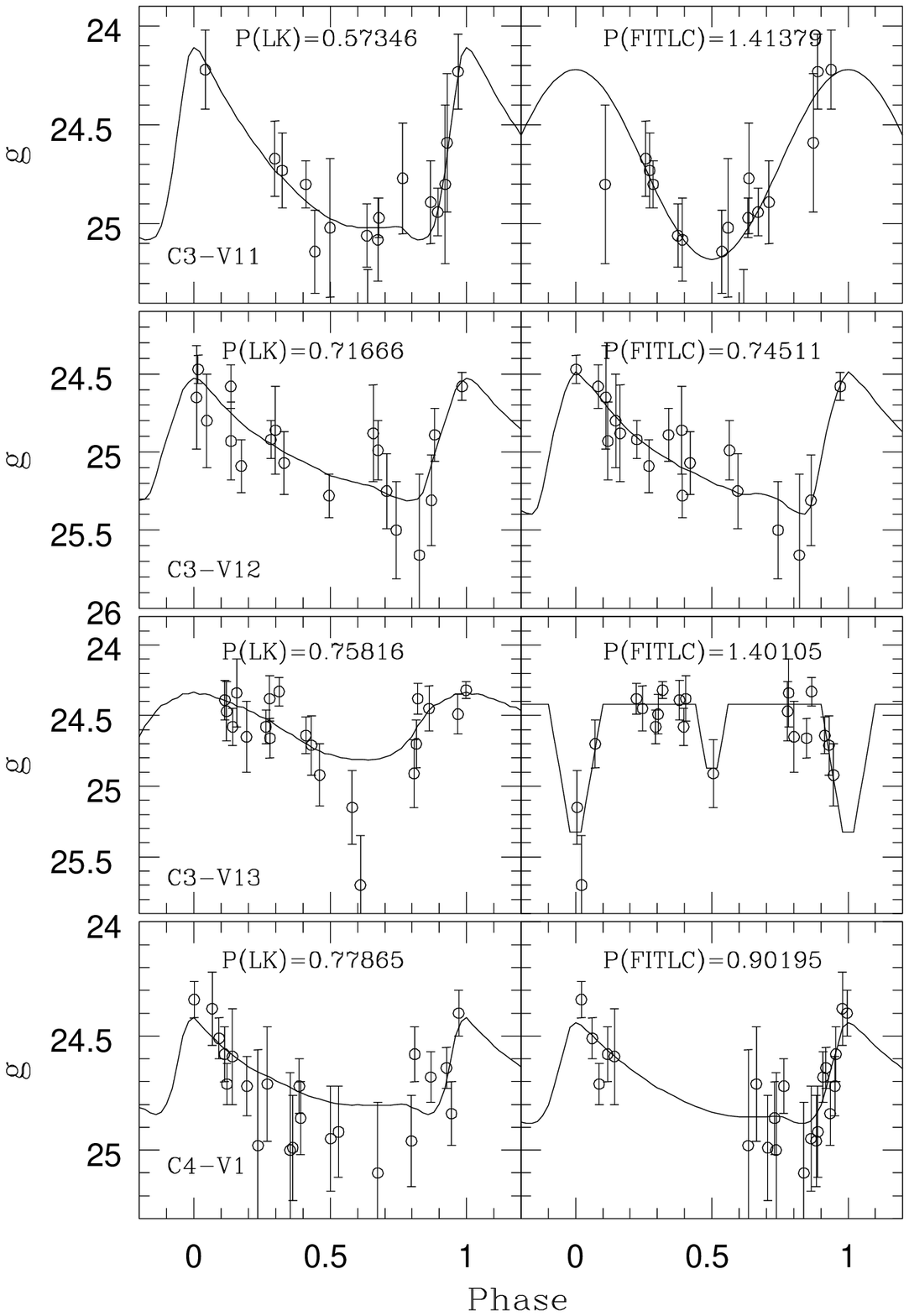}
\caption{same as Figure 3. \label{fig9)}}
\end{figure}

\clearpage

\begin{figure}
\epsscale{1.0}
\plotone{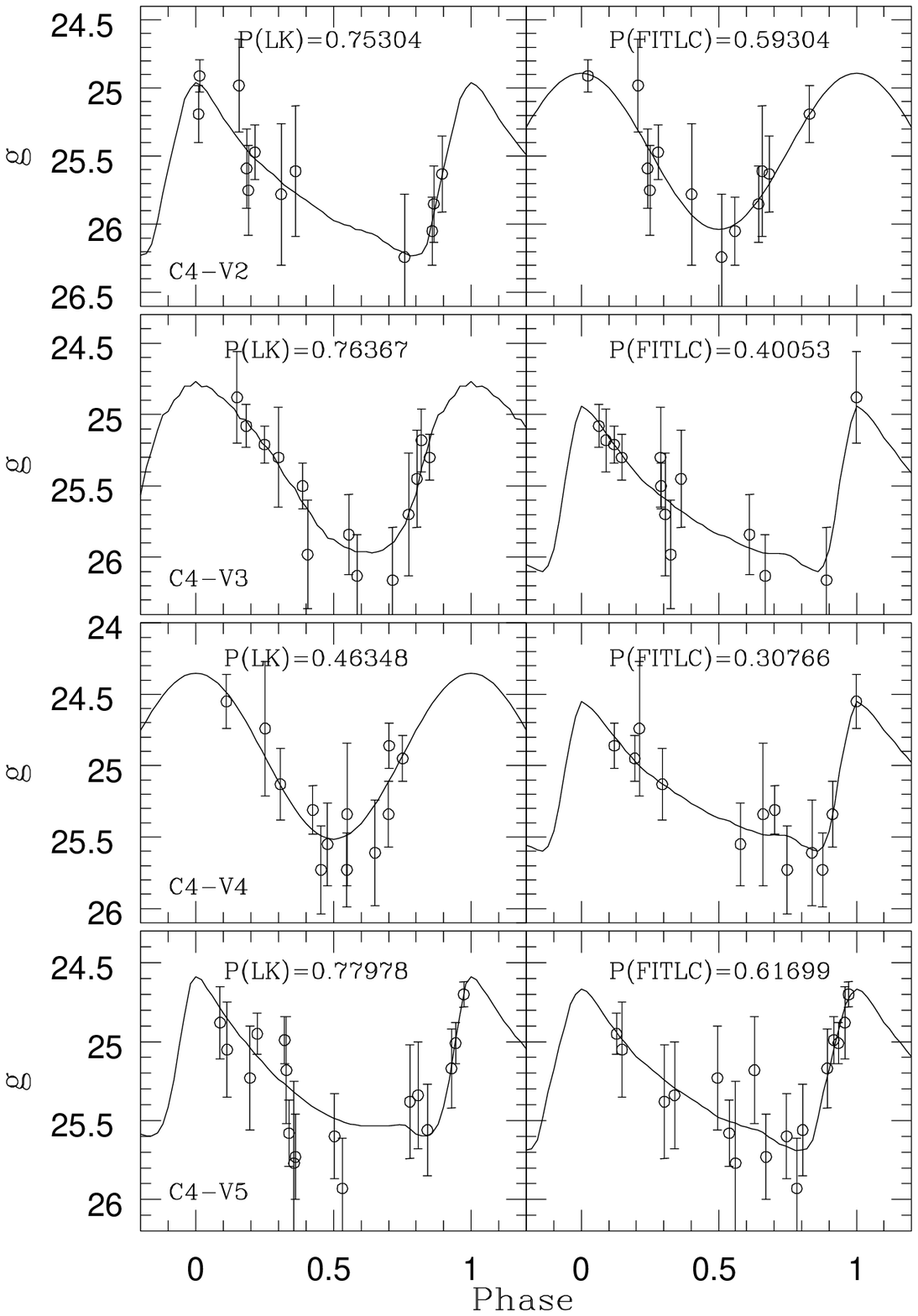}
\caption{same as Figure 3. \label{fig10)}}
\end{figure}

\clearpage

\begin{figure}
\epsscale{1.0}
\plotone{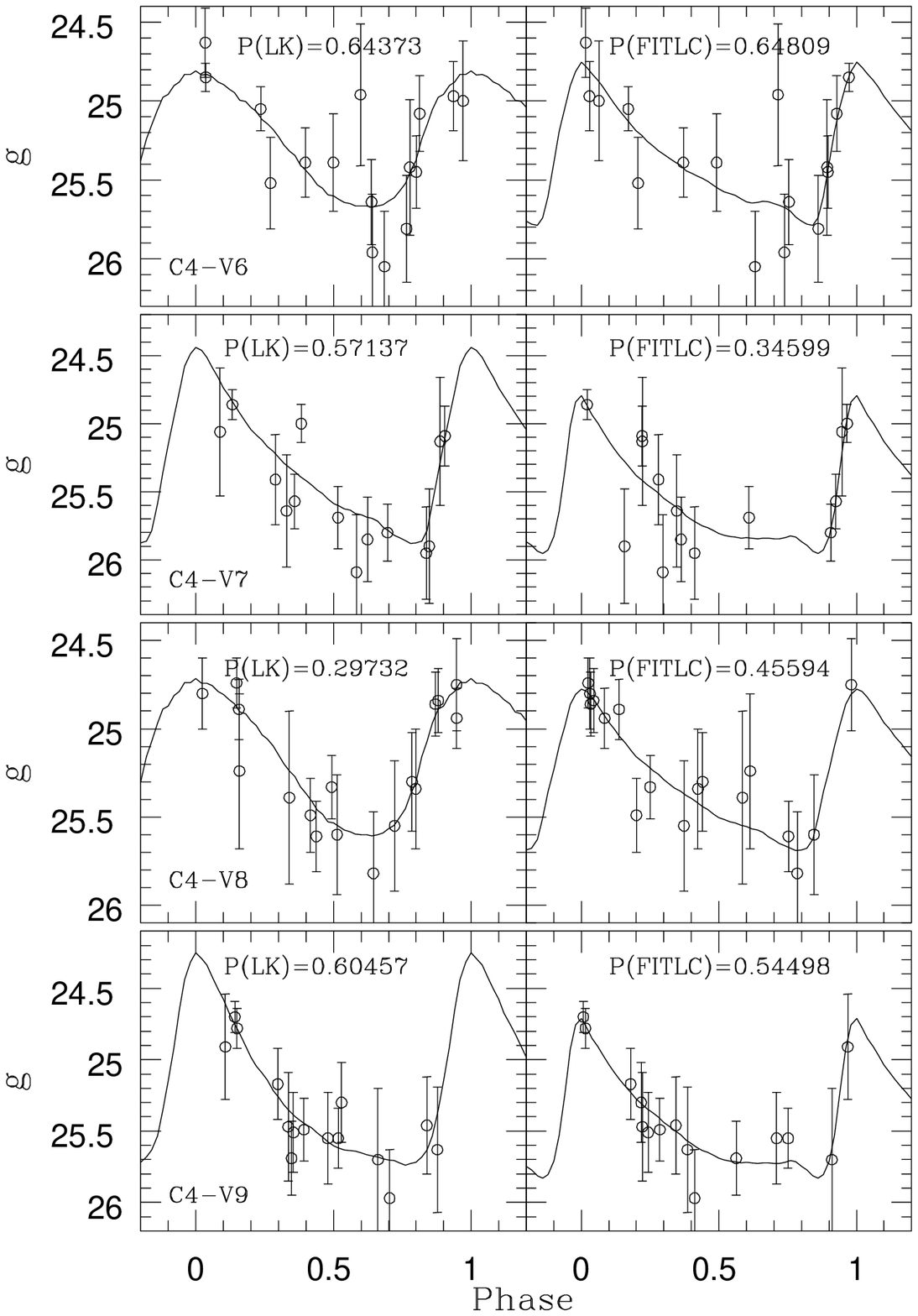}
\caption{same as Figure 3. \label{fig11)}}
\end{figure}

\clearpage

\begin{figure}
\epsscale{1.0}
\plotone{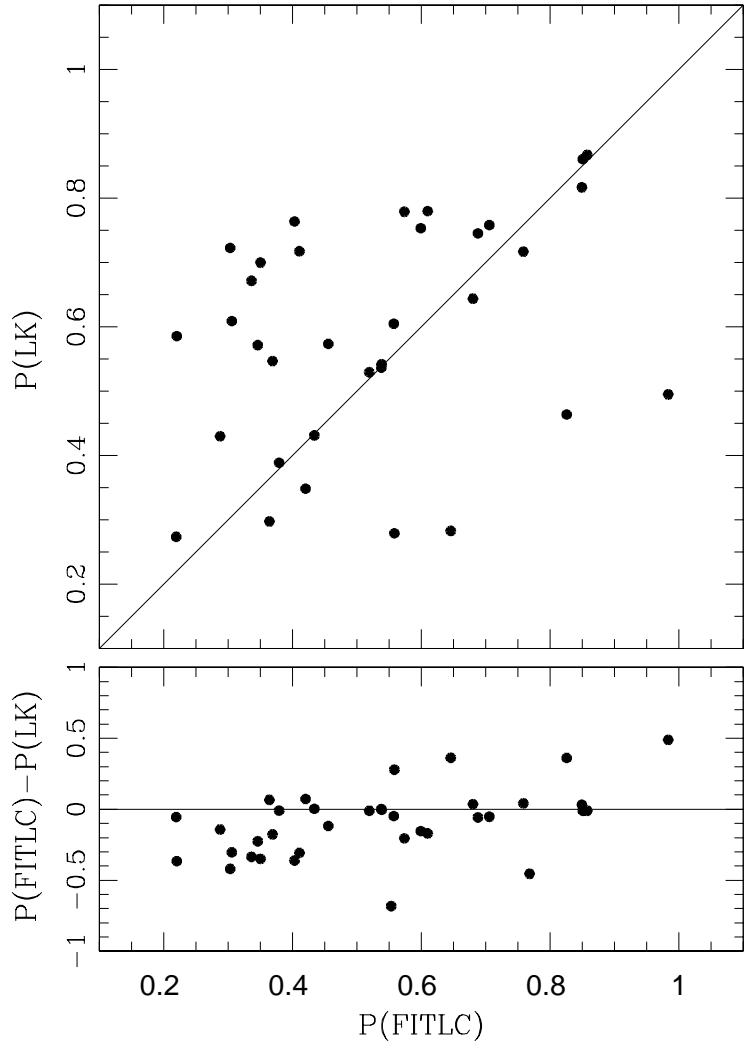}
\caption{A comparison between periods determined from the L-K and FITLC 
methods using the $g$-band photometry of RR Lyrae candidates in NGC 147. 
The L-K periods are from Saha et al (1990) while those from FITLC come from
the present study. There appears to be little correlation between the two
sets of values.  \label{fig12.}}
\end{figure}

\clearpage

\begin{figure}
\epsscale{1.0}
\plotone{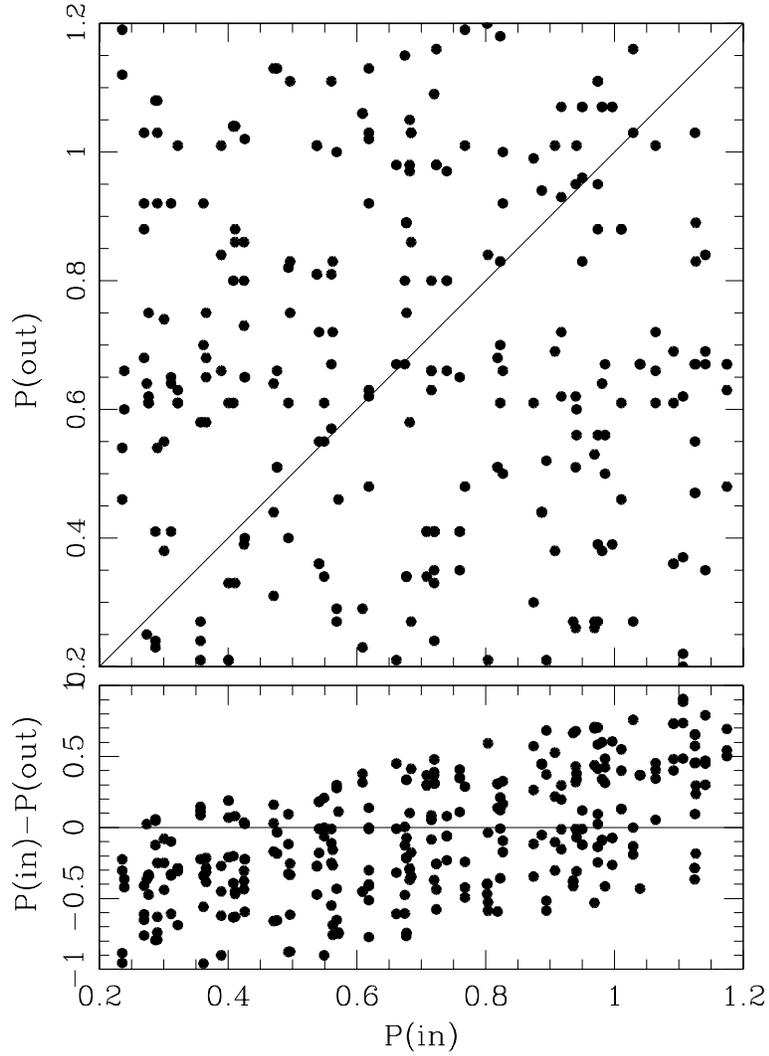}
\caption{The results of RR Lyrae variable simulations based on the
$g$-band photometry of Saha et al. (1990) with the L-K method are
shown.  The upper panel compares the input and output (recovered)
periods while the lower panel shows the difference as a function of
input period. \label{fig13}}
\end{figure}

\clearpage

\begin{figure}
\epsscale{1.0}
\plotone{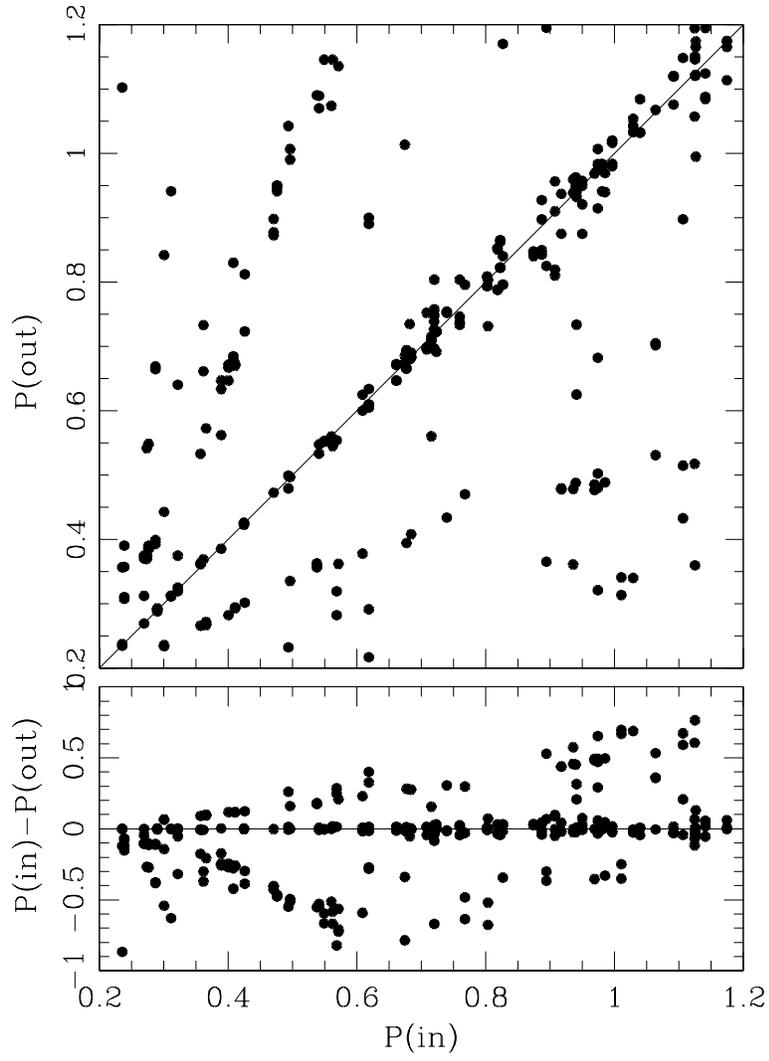}
\caption{Same as Figure 13 except that the FITLC template-fitting
algorithm is employed to determine the periods of the RR Lyrae
candidates. \label{fig14}}
\end{figure}

\clearpage

\begin{figure}
\epsscale{1.0}
\plotone{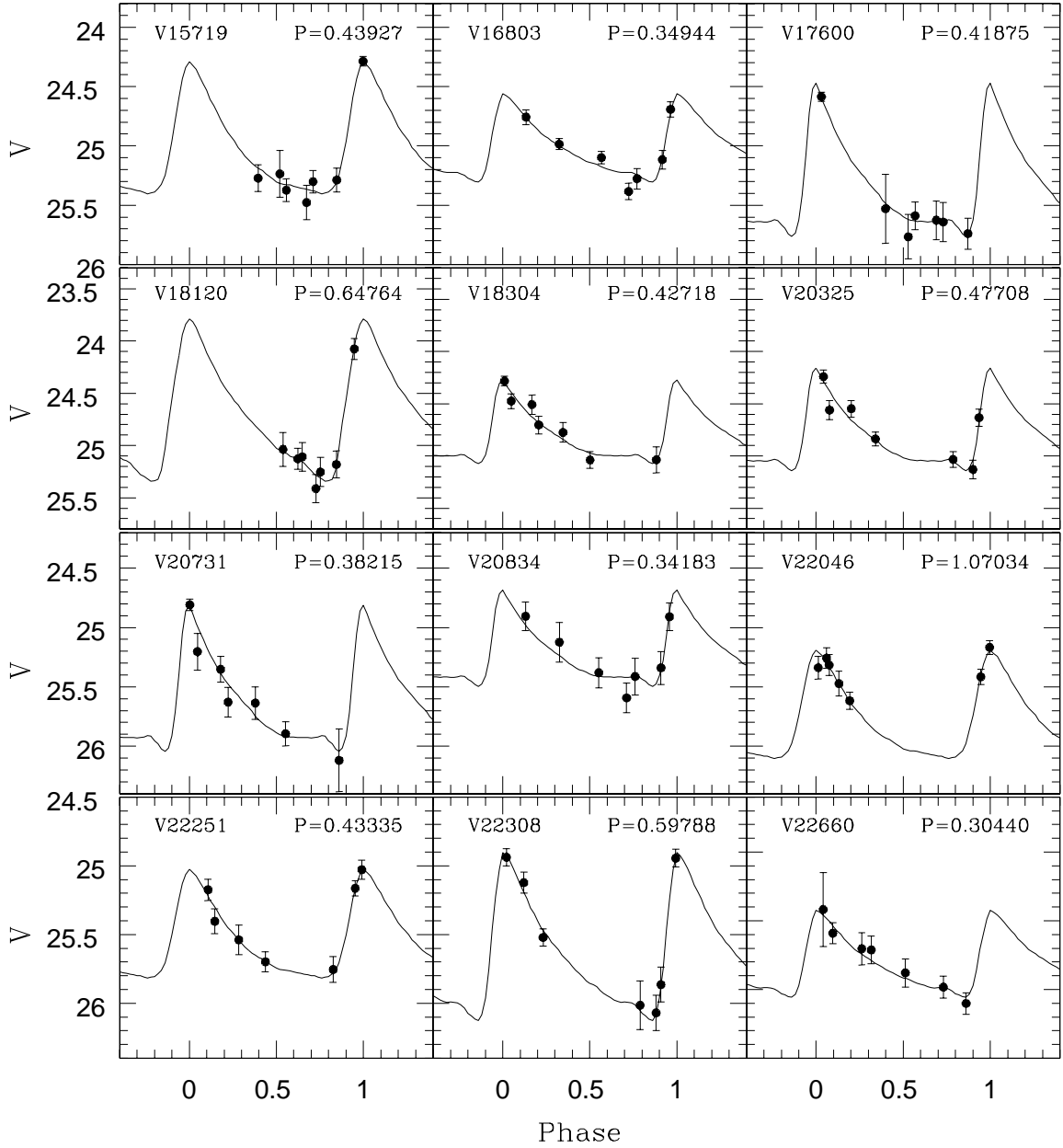}
\caption{ The best fitting template light curves for possible RR Lyrae candidates 
in NGC 147 based on archival WFPC2 imaging in the F555W filter. 
 The solid
curve is the best-fit template light curve. The star number and the
period is given in each panel. \label{fig15}}
\end{figure}

\clearpage

\begin{figure}
\epsscale{1.0}
\plotone{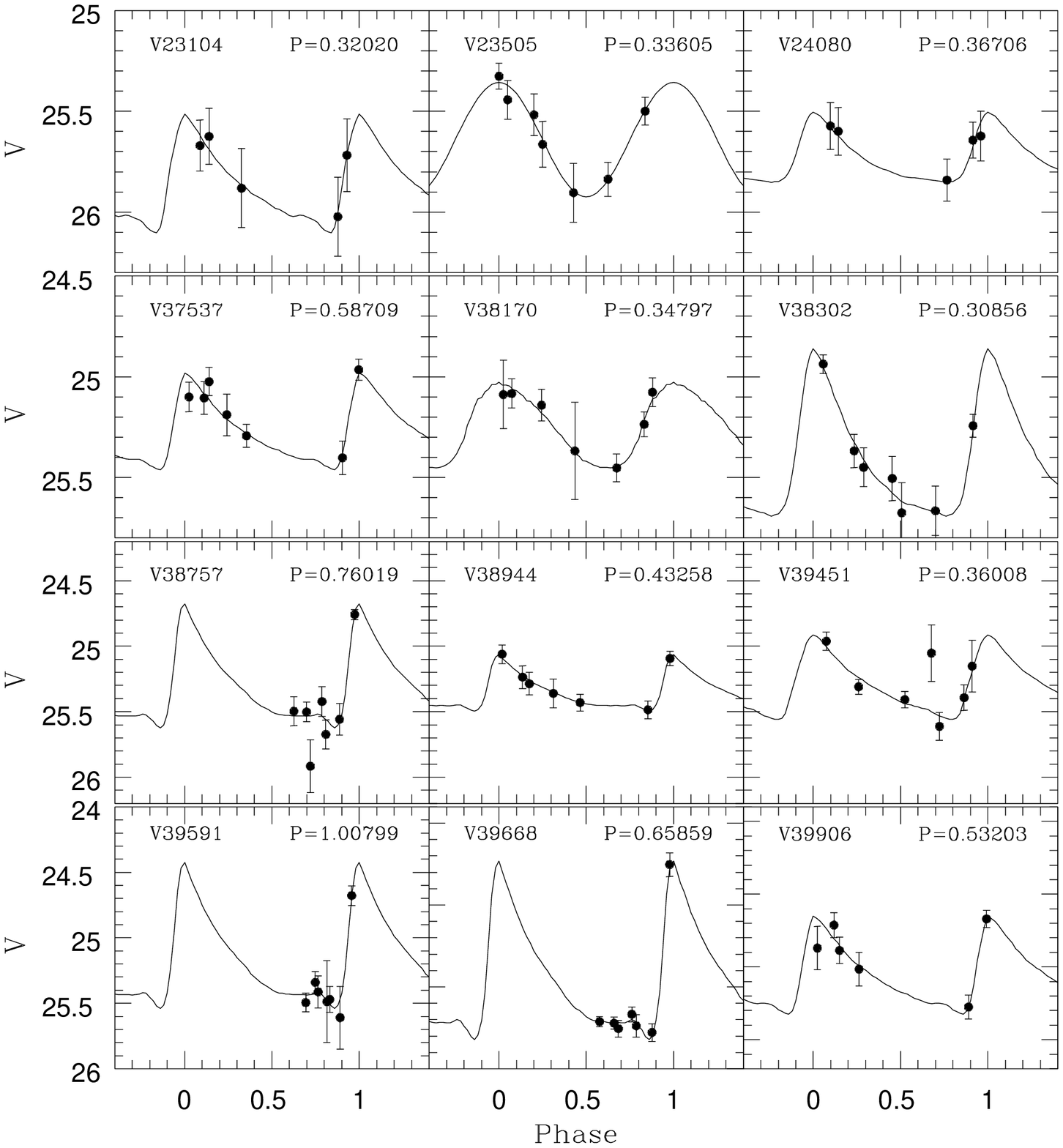}
\caption{Same as Figure 15. \label{fig16}}
\end{figure}

\clearpage

\begin{figure}
\epsscale{1.0}
\plotone{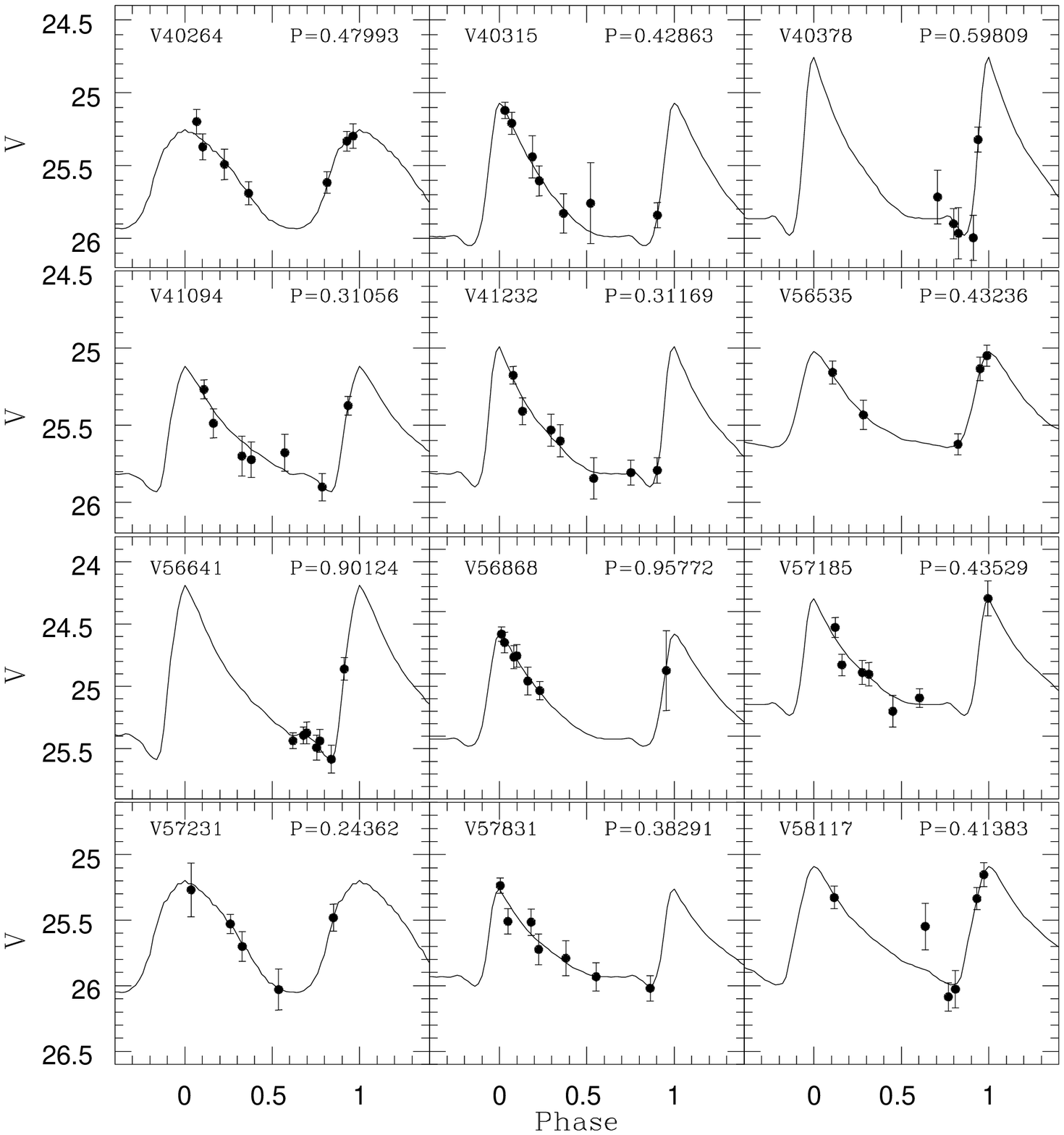}
\caption{Same as Figure 15. \label{fig17}}
\end{figure}

\clearpage

\begin{figure}
\epsscale{1.0}
\caption{The positions of the most probable RR Lyrae candidates from the 
FITLC analysis of WFPC2 archival data in the VI CMD. The open circles
are the ab-type RR Lyraes while the filled triangles are the c-type variables.
\label{fig18}}
\end{figure}

\clearpage

\begin{figure}
\epsscale{1.0}
\plotone{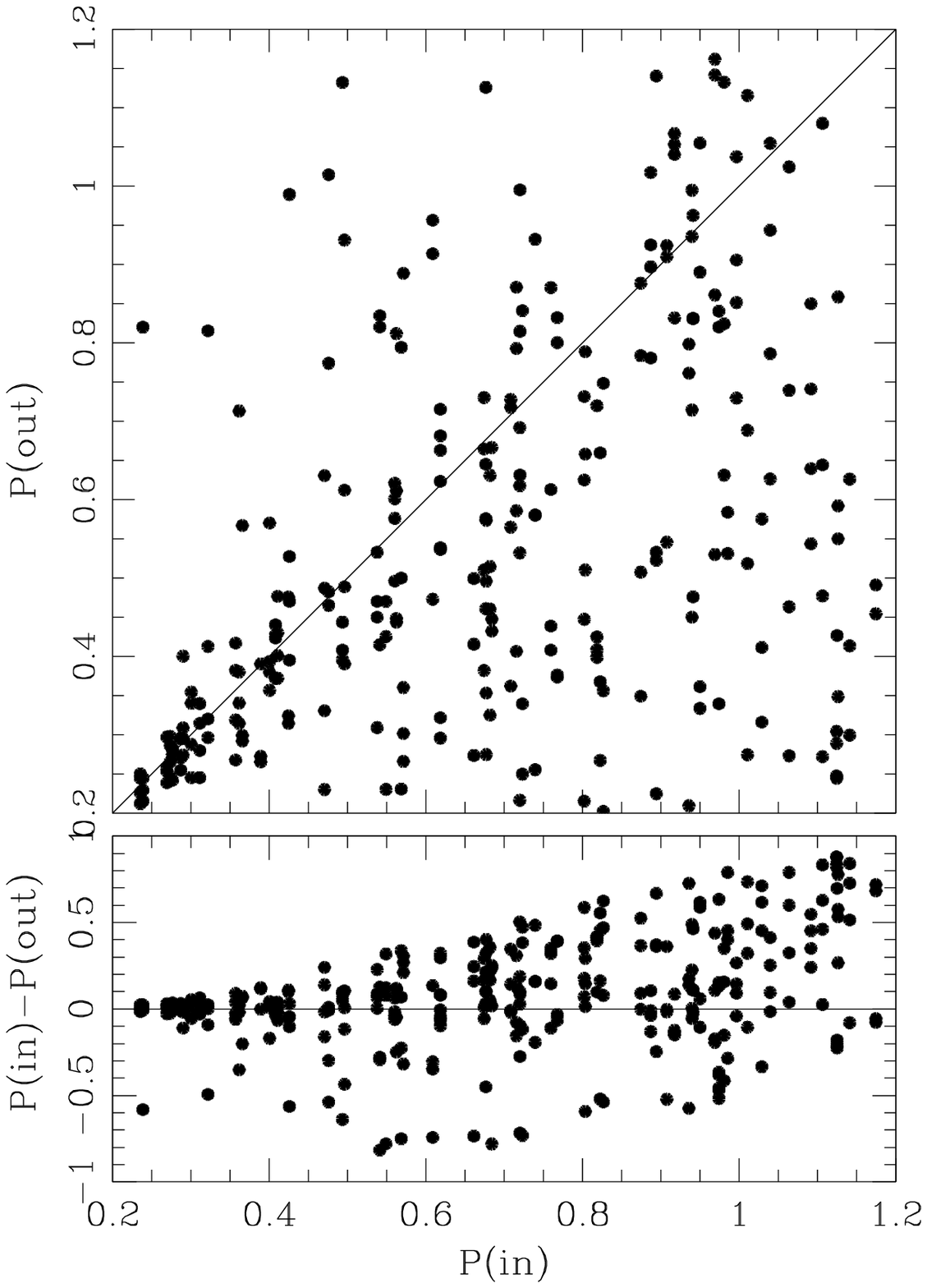}
\caption{ The result of artificial RR Lyrae variable test for WFPC2 archival data with FITLC routine. \label{fig19}}
\end{figure}

\clearpage


\begin{thebibliography}{}
\bibitem[Bender et al (1991)]{bender91} Bender, R., Paquet, A., \& Nieto, J.-L. 1991, 
\aap, 246, 349
\bibitem[]{} Bono, G., Caputo, F., Castellani, V., Marconi, M., Storm, J.
\& Degl'Innocenti, S. 2003, \mnras, 344, 1097
\bibitem[]{} Bono, G., Caputo, F., Di Criscienzo, M. 2007, \aap, 476, 779
\bibitem[Brown et al.(1978)]{brown04} Brown, T. M. et al. 2004, \aj, 127, 2738
\bibitem[Chaboyer (1999)]{chabo99} Chaboyer, B. 1999, ASSL, 237,111
\bibitem[]{} Kunder, A. \& Chaboyer, B. 2009, \aj, 138, 1284
\bibitem[Charbonneau, P. (1995)]{charb95} Charbonneau, P.  1995, \apjs, 101, 309
\bibitem[Clement, C. M. (2000)]{clem00} Clement, C. M. 2000 in IAU Colloq. 176, 266
\bibitem[Da Costa \& Mould (1988)]{daco88} Da Costa, G. S., \& Mould, J. R. 1998, \apj, 334, 159
\bibitem[Dolphin (2000)]{dolphin00} Dolphin, A. E. 2000, PASP, 112, 1383
\bibitem[]{} Fraternali, F. Tolstoy, E., Irwin, M. J., \& Cole, A. A. 2009, \aap, 499, 121
\bibitem[Grebel (1999)]{greb99} Grebel, E. K. 1999, in IAU Symp. 192, 17
\bibitem[Han et al (1997)]{han97} Han et al. 1997, \aj, 113, 1001
\bibitem[Hodge (1989)]{hodge89} Hodge, P. 1989, ARAA, 27, 139
\bibitem[Lafler \& Kinman (1965)]{lafler65} Lafler, J., \& Kinman, T. D. 1965, \apjs, 11, 216
\bibitem[Layden (1998)]{layden98}  Layden, A. 1998, \aj, 115, 193
\bibitem[Layden \& Sarajedini (2000)]{layden00} Layden, A., \& Sarajedini, A. 2000, 119,
1760
\bibitem[Mackey \& Gilmore (2003)]{mackey03}  Mackey, A. D., \& Gilmore, G. F. 2003, MNRAS, 343, 747
\bibitem[Mancone \& Sarajedini (2008)]{mancone08}  Mancone, C., \& Sarajedini, A. 2008, \aj, 136, 1913
\bibitem[Nowotny et al (2003)]{nowotny03}  Nowotny, W.,Kerschbaum, F.,Olofsson, H., \& Schwarz, H. E. 2003, A\&A,403,93
\bibitem[Saha et al (1990)]{saha90} Saha, A., Hoessel, J. G., \& Mossman, A. E. 1990, \aj, 100, 108
\bibitem[Sarajedini et al (2006)]{ata06} Sarajedini, A., Barker, M. K., Geisler, D., Harding, P., \& Schommer, R. 2006, \aj, 132, 1361
\bibitem[Schlegel et al (1998)]{schlegel98} Schlegel, D. J., Finkbeiner, D. P., \& Davis, M. 1998, \apj, 500, 525
\bibitem[Sharina et al (2006)]{sharina06} Sharina, M. E., Afanasiev, V. L., \& Puzia, T. H. 2006, \mnras, 372, 1259
\bibitem[]{} Smith, H. A. 1995, RR Lyrae Stars, Cambridge Astrophysics Series, 
ICambridge University Press; Cambridge)
\bibitem[Thuan \& Gunn (1976)]{thuan76} Thuan, T. X., \& Gunn, J. E. 1976, PASP, 88, 543
\bibitem[van den Bergh (1998)]{van98} van den Bergh 1998, \apj, 116, 1688
\bibitem[]{} Welch, D. \& Stetson, P. B. 1993, \aj, 105, 1813
\bibitem[Young \& Lo (1997)]{young97} Young, L. M., \& Lo, K. Y. 1998, \apj, 476, 127
\end{thebibliography}
\end{document}